\documentclass[pdftex,twocolumn]{aastex63}

\usepackage{apjfonts,graphicx,graphics}

\usepackage{hyperref}
\usepackage{mathptmx}
\usepackage{amsmath}
\usepackage{wasysym}
\usepackage{times}
\usepackage{epsfig}
\usepackage{ulem}
\usepackage{epstopdf}
\usepackage{appendix}

\usepackage{units}
\usepackage{import}
\newbox{\bigpicturebox}

\usepackage[capitalize]{cleveref}
\usepackage{CJKutf8}

\newcommand{\chinesenameXiaoweiduan}{{\begin{CJK}{UTF8}{gbsn}(段晓苇)\end{CJK}}}

\newcommand{\chinesenameXiaodianchen}{{\begin{CJK}{UTF8}{gbsn}(陈孝钿)\end{CJK}}}

\newcommand{\chinesenameLicaideng}{{\begin{CJK}{UTF8}{gbsn}(邓李才)\end{CJK}}}

\newcommand{\chinesenameFanyang}{{\begin{CJK}{UTF8}{gbsn}(杨帆)\end{CJK}}}

\newcommand{\chinesenameChaoliu}{{\begin{CJK}{UTF8}{gbsn}(刘超)\end{CJK}}}

\newcommand{\chinesenameHuaweizhang}{{\begin{CJK}{UTF8}{gbsn}(张华伟)\end{CJK}}}

\newcommand{\m}{\rm\; m}

\newcommand{\s}{\rm\; s}

\newcommand{\K}{\rm\; K}

\newcommand{\g}{\rm\; g}
\newcommand{\W}{\rm\; W}
\begin{document}

\title{Possible evidence of hydrogen emission in the first-overtone and multi-mode RR Lyrae variables}

\author[0000-0002-6573-6719]{Xiao-Wei Duan \chinesenameXiaoweiduan}

\affiliation{Department of Astronomy, Peking University, Yi He Yuan Road 5, Hai Dian District, Beijing 100871, China; \href{mailto:duanxw@pku.edu.cn}{duanxw@pku.edu.cn}}
\affiliation{Kavli Institute for Astronomy \& Astrophysics, Peking University, Yi He Yuan Road 5, Hai Dian District, Beijing 100871, China}
\affiliation{CAS Key Laboratory of Optical Astronomy, National Astronomical Observatories, Chinese Academy of Sciences, Beijing 100101, China; \href{mailto:licai@bao.ac.cn}{licai@bao.ac.cn}}

\author[0000-0001-7084-0484]{Xiao-Dian Chen \chinesenameXiaodianchen}
\affiliation{CAS Key Laboratory of Optical Astronomy, National Astronomical Observatories, Chinese Academy of Sciences, Beijing 100101, China; \href{mailto:licai@bao.ac.cn}{licai@bao.ac.cn}}
\affiliation{School of Astronomy and Space Science, University of the Chinese Academy of Sciences, Huairou 101408, China}
\affiliation{Department of Astronomy, China West Normal University, Nanchong 637009, China}

\author[0000-0001-9073-9914]{Li-Cai Deng \chinesenameLicaideng}
\affiliation{CAS Key Laboratory of Optical Astronomy, National Astronomical Observatories, Chinese Academy of Sciences, Beijing 100101, China; \href{mailto:licai@bao.ac.cn}{licai@bao.ac.cn}}
\affiliation{School of Astronomy and Space Science, University of the Chinese Academy of Sciences, Huairou 101408, China}
\affiliation{Department of Astronomy, Peking University, Yi He Yuan Road 5, Hai Dian District, Beijing 100871, China; \href{mailto:duanxw@pku.edu.cn}{duanxw@pku.edu.cn}}
\affiliation{Department of Astronomy, China West Normal University, Nanchong 637009, China}

\author[0000-0002-1450-9727]{Fan Yang \chinesenameFanyang}
\affiliation{CAS Key Laboratory of Optical Astronomy, National Astronomical Observatories, Chinese Academy of Sciences, Beijing 100101, China; \href{mailto:licai@bao.ac.cn}{licai@bao.ac.cn}}

\author[0000-0002-1802-6917]{Chao Liu \chinesenameChaoliu}
\affiliation{School of Astronomy and Space Science, University of the Chinese Academy of Sciences, Huairou 101408, China}
\affiliation{CAS Key Laboratory of Space Astronomy and Technology, National Astronomical Observatories, Chinese Academy of Sciences, Beijing 100101, China}

\author[0000-0001-6147-3360]{Anupam Bhardwaj}
\affiliation{Kavli Institute for Astronomy \& Astrophysics, Peking University, Yi He Yuan Road 5, Hai Dian District, Beijing 100871, China}

\author[0000-0002-7727-1699]{Hua-Wei Zhang \chinesenameHuaweizhang}
\affiliation{Department of Astronomy, Peking University, Yi He Yuan Road 5, Hai Dian District, Beijing 100871, China; \href{mailto:duanxw@pku.edu.cn}{duanxw@pku.edu.cn}}
\affiliation{Kavli Institute for Astronomy \& Astrophysics, Peking University, Yi He Yuan Road 5, Hai Dian District, Beijing 100871, China}

\begin{abstract}
The nature of shock waves in non-fundamental mode RR Lyrae stars remains a mystery because of limited spectroscopic observations. We apply a pattern recognition algorithm on spectroscopic data from SDSS and LAMOST and report the first evidence of hydrogen emission in first-overtone and multi-mode RR Lyrae stars showing the ``first apparition'', which is the most prominent observational characteristic of shock in RR Lyrae variables. We find ten RRc stars in SDSS, ten RRc stars in LAMOST, and three RRd stars in LAMOST that show blueshifted Balmer emissions. The emission features possibly indicate the existence of shock waves. We calculate the radial velocities of the emission lines, which are related to the physical conditions occurring in the radiative zone of shock waves.
Using photometric observations from ZTF, we present a detailed light curve analysis for the frequency components in one of our RRd stars with hydrogen emission, RRdl3, for possible modulations. With the enormous volume of upcoming spectral observations of variable stars, our study raises the possibility of connecting the unexplained Blazhko effect to shock waves in non-fundamental mode RR Lyrae stars.

\end{abstract}

\section{Introduction \label{sec:intro}}

The most intriguing observational property of RR Lyrae stars is that some of them show periodic amplitude and/or phase modulations, named ``Blazhko effect'' \citep{Bla1907AN....175..325B}. The light curves of these stars are modulated on timescales of tens to hundreds of days. Until now, the mechanism that causes the Blazhko effect is still under debate.
 
There have been several proposed hypotheses to explain the Blazhko effect. 
\cite{Shibahashi2000ASPC..203..299S} interpreted the Blazhko effect by an oblique-dipole magnetic rotator model. 
The lack of strong magnetic fields in the photosphere of Blazhko RR Lyrae stars disproved this explanation \citep{Chadid2004A&A...413.1087C}. 
\cite{Buchler2011ApJ...731...24B} demonstrated that irregular amplitude modulations can be triggered naturally by the nonlinear, resonant mode coupling between the fundamental mode and the 9th overtone. 
\cite{Stothers2010PASP..122..536S,Stothers2011PASP..123..127S} proposed that turbulent envelope convective cycles induce the modulations. 
\cite{Chadid2010A&A...510A..39C,Chadid2011A&A...527A.146C} firstly presented extensive photometry from space and detected significant cycle-to-cycle changes in the Blazhko modulation, which appear to be analogous to the predictions by Stothers. 

Moreover, \cite{Chadid2014AJ....148...88C} proposed that the origin of the Blazhko effect is a dynamical interaction between a multi-shock structure and an outflowing wind in a coronal structure.
The authors discovered multi-shocks propagating through the atmospheres of RR Lyrae stars through very complicated features in the RR Lyrae light curves including $jump$, $lump$, $rump$, $bump$, and $hump$, all characterized by various amplitudes and origins.  
At the Blazhko phase minimum, the Blazhko amplitude is at the minimum, the rise time is at the maximum, the shock intensity is smaller, the amplitude of the bump is lower, and the star roughly looks like an RRc. Whereas at the Blazhko phase maximum, the Blazhko amplitude is at the maximum, the rise time is at the minimum, the shock intensity is higher, the amplitude of the bump is stronger, and the light curve of the Blazhko star is similar to an RRab.

In order to study the dynamics of shock waves in the atmosphere of RR Lyrae, we can trace the observational characteristics of the spectra. 
Propagation of shock waves through the stellar atmosphere leads to hydrogen emission lines, helium emission lines, line broadening and doubling phenomena, and neutral metallic line disappearance phenomena in RR Lyrae stars \citep{Iroshnikov1962SvA.....5..475I,Chadid2008A&A...491..537C,Chadid2017ApJ...835..187C}. However, these phenomena were not found in all types of RR Lyrae. 
RR Lyrae stars can be classified by their pulsations as fundamental mode (RRab), first overtone (RRc), or multi-mode (RRd) RR Lyrae according to the number of oscillation modes \citep{Soszynski2016MNRAS.463.1332S,Beaton2018SSRv..214..113B,Bhardwaj2020JApA...41...23B}.
As for RRab stars, three moderate-to-weak hydrogen emission lines, named ``three apparitions'', appear sequentially in the spectra during a pulsation cycle \citep{Preston2011AJ....141....6P}. 
The ``first apparition'' is a blueshifted emission line, generated when the shock wave is close to the photosphere, indicating the coming of the maximum luminosity. The ``second apparition'' is a weak hump at the blue emission shoulder of H$\alpha$, which appears during the pre-minimum brightening near $\phi=0.7$. This weak hump is thought to be produced by photospheric compression caused by the inner deep expanding atmosphere colliding with the outer layers which are experiencing a ballistic infalling motion \citep{Hill1972}. The ``third apparition'' is a weak redshifted emission that appears during declining light near $\phi=0.3$ \citep{Chadid2013MNRAS.434..552C}.

The signatures of shock waves can be routinely detected in RRab stars, but no detections have been made in RRc or RRd stars. 
The link between the Blazhko effect and shock waves is unknown due to this absence.  
In this work, we develop a pattern searching algorithm, and apply it to large spectroscopic databases, in order to investigate shock waves in RR Lyrae stars, especially in non-fundamental mode variables.  
We report the first evidence of hydrogen emission in first-overtone and multi-mode RR Lyrae stars from low-resolution spectroscopic surveys, which can provide new insights into the nature of non-fundamental mode RR Lyrae stars.

In the following, we describe the structure of this article. In Section~\ref{sec:obr}, we describe the observations and methods, including our pattern recognition pipeline. In Section~\ref{section:results}, we present the search and measurement results. Then we further discuss how to understand the ``first apparition'' and Blazhko effect in non-fundamental RR Lyrae in Section~\ref{section:discussion}. Moreover, we elaborate on a detailed analysis of the full light curve solution and a possible modulation for one of the selected RRd stars in Section~\ref{section:RRdl3}. Finally, our conclusions are covered in Section~\ref{section:conclusion}.

\section{Observations and methods} \label{sec:obr}

To investigate shock waves in RR Lyrae stars, we need both photometric and spectroscopic observations. Time-domain photometry was used to identify the types of RR Lyrae stars, which was retrieved from the open resources of the Catalina Sky Survey \citep{Drake2014,Drake2017}, Gaia DR2 \citep{Clementini2019A&A...622A..60C}, Wide-field Infrared Survey Explorer \citep[WISE,][] {chen2018}, All-Sky Automated Survey for Supernovae \citep[ASAS-SN,][]{Jayasinghe2019MNRAS.486.1907J}, and Asteroid Terrestrial-impact Last Alert System \citep[ATLAS,][]{Heinze2018AJ....156..241H}. After combining these catalogs and removing duplicate data we get 68,152 RR Lyrae stars. We also get light curves from the Zwicky Transient Facility \citep[ZTF,][]{Bellm2019PASP..131a8002B,Chen2020} for our selected stars if available.

We search for the ``first apparitions'' among low-resolution and single-epoch spectra from the Sloan Digital Sky Survey \citep[SDSS,][]{Eisenstein2011AJ....142...72E} and the Large Sky Area Multi-Object Fiber Spectroscopic Telescope survey \citep[LAMOST,][]{Deng2012RAA....12..735D}. Due to the fact that periods of RR Lyrae stars are rather short, and the ``first apparition'' only enjoys about $5\%$ of the whole pulsation cycle \citep{Chadid2011Carnegie}, long-time exposures and co-addition of spectra smooth out the relevant emission lines. In both SDSS and LAMOST surveys, each plate (sky pointing) was observed three times, 10$-$20 minutes each for SDSS and 25 minutes each for LAMOST. This gives the best possible time resolution for variable targets, which turns out to be good enough for our purpose of studying RR Lyrae stars. 

Some of the SDSS spectra come from BOSS. Spectra of SDSS cover wavelength from 3,800 to 9,200 $\rm \mathring{A}$ and those of BOSS have wavelength coverage of 3,650$-$10,400 $\rm \mathring{A}$. Both data provide resolutions of 1,500 at 3,800 $\rm \mathring{A}$ and 2,500 at 9,000 $\rm \mathring{A}$. We choose ``spCFrame'' spectra, which achieve nominal resolutions of about $4,150$ at H$\alpha$ and $4,180$ at H$\beta$. In the dataset of LAMOST DR6, the spectral coverage is from 3,550$-$9,110 $\rm \mathring{A}$ with $R \sim 1,800$, the nominal resolutions are about $7,800$ at H$\alpha$ and $8,400$ at H$\beta$. After pair matching the catalogs  using TOPCAT \citep{Taylor2005ASPC..347...29T}, we retrieve RR Lyrae stars 3,526 from SDSS DR7$-$9 and 5,571 from LAMOST DR1$-$6. All selected targets have both light curves and spectra.

To hunt for the ``first apparitions'', the common practice is to visually check the profiles of hydrogen Balmer lines for any sign of emission in a small sample. For large survey data, this can become an enormous amount of work. 
Therefore, we build a pipeline using a handcrafted one-dimensional pattern recognition method. We presuppose the pattern of the ``first apparition'' as a Gaussian-like emission profile on the blue wing of a broad Gaussian-like absorption profile when both two signals are more significant than $2\sigma$ compared to the average level. We assume that the minimum flux value in the selected windows (e.g., 6,540-6,590 $\rm \mathring{A}$ for H$\alpha$) indicates the location of the Balmer absorption profile. We retain the results of the hunt that show clear patterns of the ``first apparition'' in H$\alpha$ and H$\beta$ simultaneously and contain at least two observational points on the profiles of emission. We apply the pipeline to the spectra of RR Lyrae stars from SDSS and LAMOST, with visual inspection to check false negative.

The components of emission and absorption are fitted based on the $scale$ $width$ $versus$ $shape$ method for the Balmer lines \citep{Sersic1968adga.book.....S,Clewley2002MNRAS.337...87C}. We adopt two $S\acute{e}rsic$ profiles \citep{Xue2008,Yang2014} as:
\begin{eqnarray}\label{eq:Sersicprofile}
y =m - a e^{-(\frac{\left|{\lambda}-{\lambda_0}\right|}{b})^c},
\end{eqnarray}
to fit the profile and measure the flux, intensity and full width at half maximum (FWHM) for the two components. Uncertainties are generated by error propagation using the covariance matrix and Monte Carlo method \citep{Andrae2010arXiv1009.2755A}.

\section{Results}\label{section:results}

\begin{figure}[tb]
\begin{center}
\begin{tabular}{c}
\includegraphics*[width=90.3mm,height=6cm]{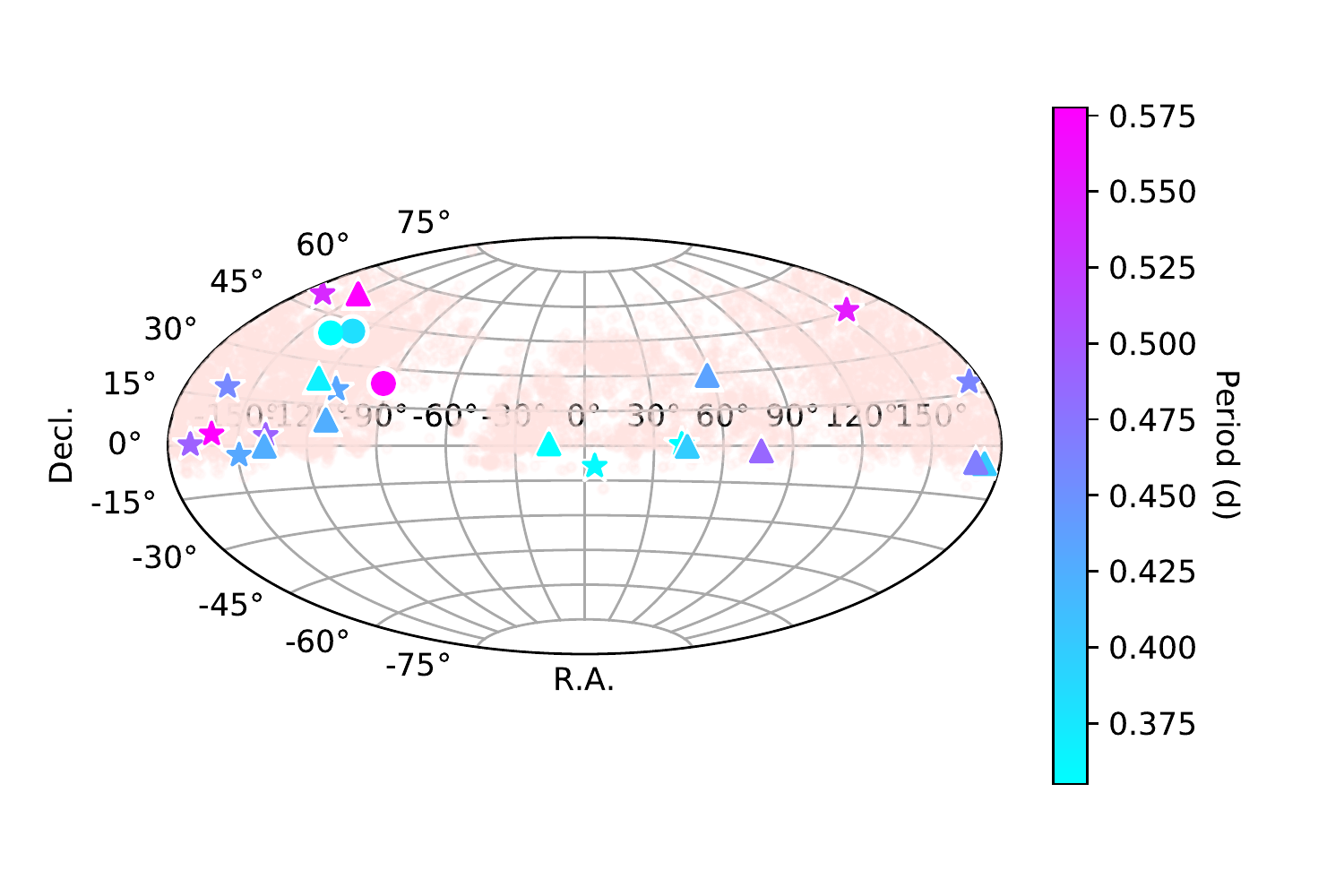}
\end{tabular}
\caption{\label{fig:spatialdistribution} Visualization of the spatial distribution of RRc and RRd stars from the SDSS and LAMOST datasets. Stars with a detected ``first apparitions'' are highlighted with five-pointed stars (RRc in SDSS), triangles (RRc in LAMOST) and circular points (RRd in LAMOST), while light pink points in the background show the whole RRc and RRd sample from which our survey is based. The variation of periods is indicated by different colors.}
\end{center}
\end{figure}

\begin{figure*}[tb]
\begin{center}
\begin{tabular}{c}
\includegraphics*[width=180mm,height=14.4cm]{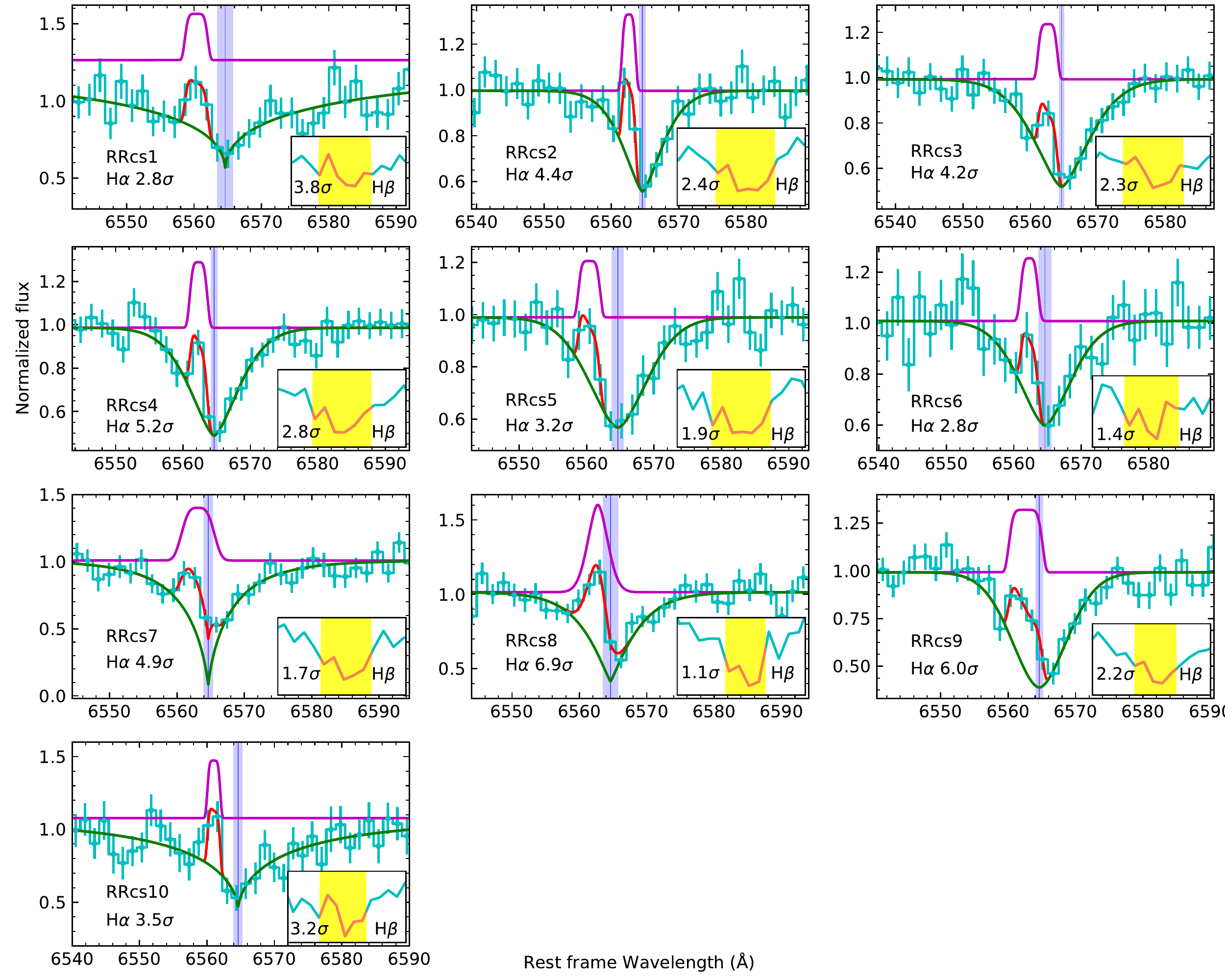}
\end{tabular}
\caption{\label{fig:SDSSexample}Visualization of fitting results of RRc that show the ``first apparitions'' in SDSS.  
The reference frame for the wavelength axis is the stellar rest frame. The blueshifted H$\alpha$ emission lines are displayed as pink profiles. Vertical blue lines indicate the H$\alpha$ line laboratory wavelength. The H$\beta$ profiles are displayed in the subplots. The significances of the emission lines indicate the signal-to-noise ratio.}\end{center}
\end{figure*}

\begin{figure*}[tb]
\begin{center}
\begin{tabular}{c}
\includegraphics*[width=180mm,height=21.6cm]{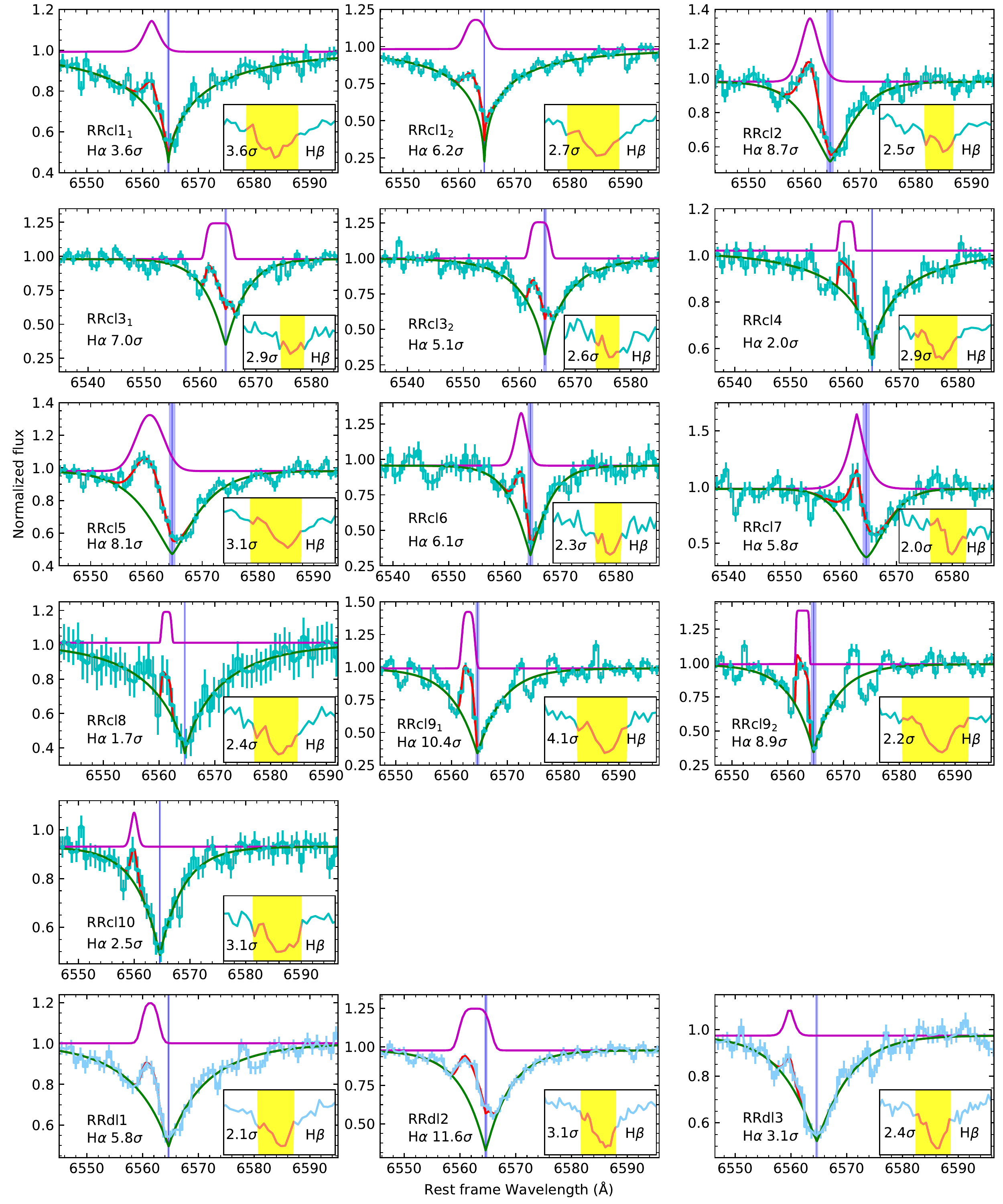}
\end{tabular}
\caption{\label{fig:LAMOSTexample}
As Figure~\ref{fig:SDSSexample}, but for RRc and RRd stars that show shock wave observational characteristics in LAMOST.}\end{center}
\end{figure*}

As one of the main results of the search, we find ten RRc stars in SDSS, ten RRc stars in LAMOST, and three RRd stars in LAMOST that show shifted emission components on the blue wing of the Balmer absorption lines. 
The spatial distribution of those RR Lyrae is shown in Figure~\ref{fig:spatialdistribution}.  
 Figures~\ref{fig:SDSSexample} and \ref{fig:LAMOSTexample} show the SDSS and LAMOST spectra respectively, which display the feature in the H$\alpha$ line. 
 The reference frame for the wavelength axis is the stellar rest frame. The flux is normalized by the continuum. The blueshifted H$\alpha$ emission lines are shown as pink profiles. The H$\beta$ profiles are displayed in the subplots. We calculate the significance of the emission lines as the signal-to-noise ratio. We choose the candidates with signals of H$\alpha$ exceeding $2\sigma$ and the signals of H$\beta$ exceeding $1\sigma$. 

\begin{figure*}[tb]
\begin{center}
\begin{tabular}{c}
\includegraphics*[width=180mm,height=6cm]{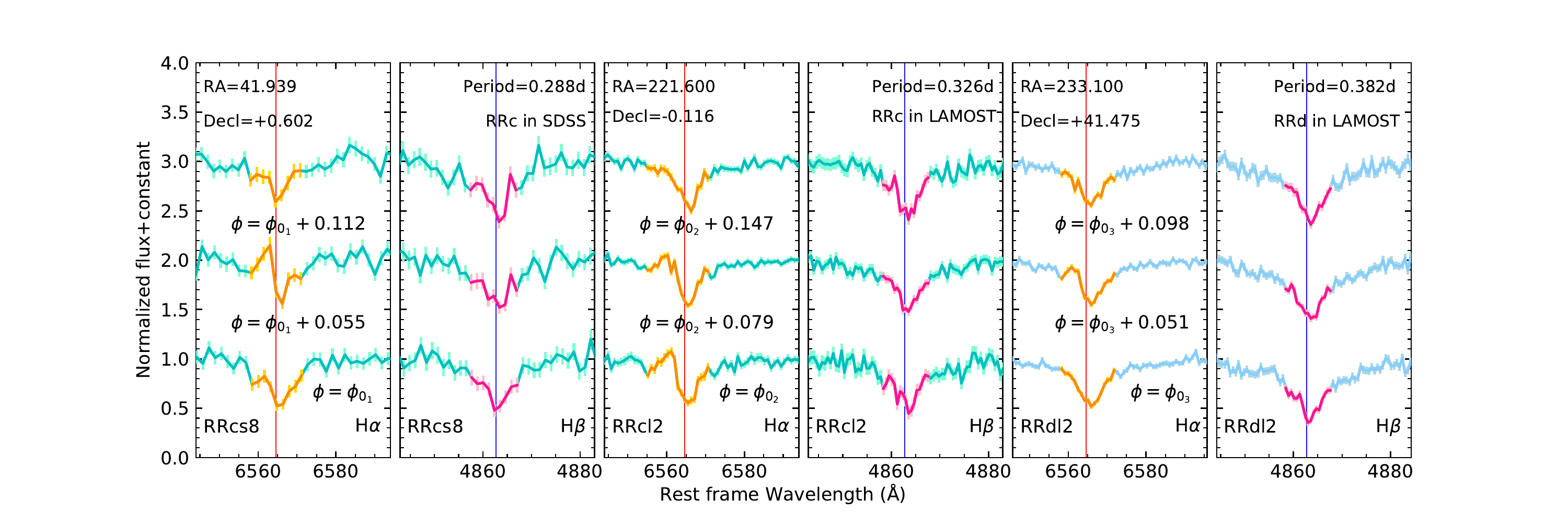}
\end{tabular}
\caption{\label{fig:evolutionexample} Evolution of the H$\alpha$ and H$\beta$ line profiles of RRcs8, RRcl2, and RRdl2. The frame of reference for the wavelength axis is the stellar rest frame. The vertical red lines indicate the H$\alpha$ line laboratory wavelength, while the vertical blue lines indicate the H$\beta$ line laboratory wavelength.}
\end{center}
\end{figure*}

The significances of the emission lines certify their physical reality. 
Moreover, the probability that the spectra show significant signals at H$\alpha$ and H$\beta$ at the same time due to the statistical effects of noise is close to zero. 
The correlation between emission and absorption features appear to distort the location of their centroids. There are slight differences between the redshifts given by the centroids of the absorption lines and given by the SEGUE Stellar Parameter Pipeline \citep[SSPP,][]{Lee2008AJ....136.2022L} or the LAMOST stellar parameter pipeline \citep[LASP,][]{Wu2014IAUS..306..340W} in some stars. We display examples of the evolution of the H$\alpha$ and H$\beta$ line profiles of RRc in SDSS and LAMOST and RRd in LAMOST in Figure~\ref{fig:evolutionexample}. To avoid the miscalculation of the phase due to the Blazhko effect, we only present the differences of phases. The emission lines are gradually getting intenser or weaker, which indicates that the signals are real evolving emissions instead of coincidences due to noise.

Three RRc stars in LAMOST show blueshifted Balmer emissions in multiple spectra of a single plane, whereas no such candidates are found in SDSS, possibly due to our strict selection criteria. 
We also get no RRd stars with ``first apparitions'' in SDSS. We find two RRd stars in SDSS that may contain the emission feature, but we reject them because there is only one observational point on the emission profile.

Parameters of the stars and measurements of the emission lines are summarized in Table~\ref{table:parameter}, and \ref{table:measurement}.
We collect parameters generated by SSPP, LASP, and \cite{Liu2020arXiv200201188L}. As for SSPP and LASP, these two pipelines also report primary stellar parameters such as effective temperature ($T_{\rm eff}$), surface gravity (log $g$), and their uncertainties for most stars in the temperature range 4000$-$10,000 K, with spectral S/N ratios exceeding 10. \cite{Liu2020arXiv200201188L} adopted a new method to measure the metallicities of RR Lyrae stars. Considering the nature of pulsating stars, they provided measurements for individual single-epoch spectra and adopted mean values weighted by S/N as the final metallicity to avoid the influence from the ``apparitions''. They fixed $\alpha-$element to iron abundance ratio, [$\alpha$/Fe], to 0.4, which makes their measurement result more sensitive for metal-poor stars than LASP, but suffer from larger uncertainties than SSPP, so we adopt it for the LAMOST dataset. 
For the emission component of the ``first apparition'', we also measure the redshift in the frame of the observer, radial velocity in the stellar rest frame, normalized flux, and FWHM.

\begin{table*}[tbp]
\caption{\label{table:parameter} Parameters of selected RRc and RRd stars from SDSS and LAMOST sample.}
\begin{center}
\setlength{\tabcolsep}{1mm}{
\begin{tabular}{c}
RRc stars from SDSS sample \\
\end{tabular}
\begin{tabular}{lccccccccccccccc}
\hline
{Object}   & R.A.(J2000) & Decl.(J2000) & Period & $V$ & Amp & $z^s$ & $T_{\rm eff}^S$ & log $g^S$ & [Fe/H]\\
   $  $          & $({}^{\circ})$ & $({}^{\circ})$ & day & mag & mag & $ $ & K & $ $ & $  $ \\
\hline
RRcs1  & $198.35858$ & $+18.16208$ & $0.33851$ & $16.78$ & $0.42$ & $ -2.05E-4\pm1.69E-5$ & $7626.91\pm99.38$ & $3.56\pm0.37$ & $-1.79\pm0.05$ \\
RRcs2  & $174.79487$ & $+18.36381$ & $0.34043$ & $16.88$ & $0.35$ & $ 2.17E-4\pm1.49E-5$ & $6783.39\pm78.43$ & $2.82\pm0.18$ & $-1.69\pm0.002$ \\
RRcs3  & $210.54617$ & $-3.02092$ & $0.32518$ & $17.31$ & $0.53$ & $ 4.95E-4\pm1.11E-5$ & $7443.48\pm138.66$ & $3.71\pm0.26$ & $-1.63\pm0.16$ \\
RRcs4  & $221.99438$ & $+3.47414$ & $0.35448$ & $17.37$ & $0.36$ & $ -6.13E-4\pm1.19E-5$ & $7394.68\pm38.43$ & $3.28\pm0.40$ & $-1.61\pm0.06$ \\
RRcs5  & $246.94200$ & $+20.78986$ & $0.32556$ & $17.41$ & $0.36$ & $ -5.53E-4\pm2.22E-5$ & $7065.13\pm58.53$ & $3.05\pm0.23$ & $-1.26\pm0.07$ \\
RRcs6  & $189.85158$ & $+49.18964$ & $0.37950$ & $17.55$ & $0.43$ & $ 1.32E-4\pm3.06E-5$ & $6978.94\pm158.95$ & $3.30\pm0.68$ & $-2.33\pm0.01$ \\
RRcs7  & $155.54754$ & $+45.33461$ & $0.38524$ & $17.62$ & $0.37$ & $ -5.12E-4\pm1.43E-5$ & $6982.75\pm86.32$ & $2.94\pm0.16$ & $-2.06\pm0.03$ \\
RRcs8  & $41.93854$ & $+0.60214$ & $0.28773$ & $17.73$ & $0.47$ & $ -5.56E-4\pm1.92E-5$ & $7241.95\pm56.80$ & $3.25\pm0.36$ & $-1.57\pm0.08$ \\
RRcs9  & $198.54504$ & $+3.55614$ & $0.39695$ & $17.97$ & $0.48$ & $ 6.44E-5\pm1.40E-5$ & $7245.77\pm98.90$ & $3.05\pm0.52$ & $-2.51\pm0.10$ \\
RRcs10  & $189.62392$ & $+0.33389$ & $0.35529$ & $18.63$ & $0.36$ & $ 4.96E-5\pm1.20E-5$ & $/$ & $/$ & $-1.92\pm0.23^{*}$ \\
\hline
\end{tabular}
\begin{tabular}{c}
RRc and RRd stars from LAMOST sample \\
\end{tabular}
\begin{tabular}{lccccccccccccccc}
\hline
{Object}   & R.A.(J2000) & Decl.(J2000) & Period & $V$ & Amp & $z^L$ & [Fe/H] \\
   $  $          & $({}^{\circ})$ & $({}^{\circ})$ & day & mag & mag & $ $ & $ $  \\
\hline
RRcl1  & $247.04771$ & $9.45261$ & $0.32534$ & $14.91$ & $0.41$ & $ -8.53E-4\pm1.65E-6$ & $-1.41\pm0.09^{*}$ \\
RRcl2  & $221.59958$ & $-0.11594$ & $0.32560$ & $14.92$ & $0.35$ & $ -4.30E-4\pm2.27E-5$ &  $-1.57\pm0.10^{*}$ \\
RRcl3  & $169.58008$ & $-4.83961$ & $0.35061$ & $15.41$ & $0.42$ & $ 9.39E-4\pm6.88E-5$ & $ / $ \\
RRcl4  & $58.24550$ & $+29.20903$ & $0.33122$ & $15.53$ & $0.35$ & $ 3.80E-4\pm5.37E-5$ &  $-1.68\pm0.14^{*}$ \\
RRcl5  & $344.53867$ & $+0.87136$ & $0.28274$ & $15.82$ & $0.39$ & $ -4.93E-4\pm5.60E-5$ &  $-1.30\pm0.14^{*}$ \\
RRcl6  & $173.43429$ & $-5.10247$ & $0.31021$ & $15.86$ & $0.36$ & $ 4.39E-4\pm3.10E-5$ & $-1.65\pm0.25^{*}$ \\
RRcl7  & $76.40079$ & $-2.05489$ & $0.36290$ & $16.26$ & $0.32$ & $ 6.22E-4\pm6.00E-6$ & $ / $ \\
RRcl8  & $44.33796$ & $-0.22086$ & $0.30943$ & $16.77$ & $0.30$ & $ -1.96E-4\pm1.16E-5$ & $-1.29\pm0.25^{*}$ \\
RRcl9  & $206.81642$ & $+52.56544$ & $0.41721$ & $16.81$ & $0.41$ & $ -9.41E-4\pm4.98E-6$ & $ / $ \\
RRcl10  & $236.50104$ & $+24.13336$ & $0.29025$ & $17.03$ & $0.33$ & $ -1.02E-3\pm7.48E-5$ & $ / $ \\
\hline
RRdl1  & $266.95863$ & $+24.17853$ & $0.57764$ & $14.93$ & $0.77$ & $ -8.12E-4\pm1.91E-5$ & $-2.18\pm0.09^{*}$\\
RRdl2  & $233.10042$ & $+41.47511$ & $0.38228$ & $15.61$ & $0.39$ & $ -6.83E-4\pm2.12E-5$ & $ / $ \\
RRdl3  & $223.47962$ & $+39.53931$ & $0.35499$ & $15.81$ & $0.33$ & $ -9.47E-4\pm8.73E-6$ & $-1.50\pm0.20^{*}$\\
\hline
\hline
\end{tabular}}
\end{center}
\tablecomments{\\\hspace{\textwidth}
 1. Period, $V$ and Amp (Amplitude) are produced by Catalina Sky Survey.\\\hspace{\textwidth}
 2. $z^s$, $T_{\rm eff}^S$, log $g^S$ and [Fe/H] (without $*$) are generated from SSPP. $z^L$ are generated from LASP.\\\hspace{\textwidth}
 3. [Fe/H] (with $*$) are given by \cite{Liu2020arXiv200201188L}. [Fe/H] of RRcs10 is lack of data in SSPP.\\\hspace{\textwidth}
 4. Unavailable values are marked using a backslash $/$.}
\end{table*}

\begin{table*}[tbp]
\caption{\label{table:measurement} Measurements of selected RRc and RRd stars from SDSS and LAMOST sample.}
\begin{center}
\begin{tabular}{lcccc}
\hline
{Object}  & $z_{\rm e1,\alpha}$ & $V_{\rm e1,\alpha}$ &  Flux$_{\rm e1,\alpha}$ & FWHM$_{\rm e1,\alpha}$\\
   $  $     & $ $ & km/s &   & $\rm \mathring{A}$   \\
\hline
RRcs1  & $-9.44E-4\pm1.87E-4$ & $-195\pm79$ &  $0.30$ & $3.33$\\
RRcs2  & $-1.97E-4\pm1.88E-4$ & $-93\pm61$ &  $0.33$ & $2.02$\\
RRcs3  & $1.13E-4\pm1.88E-4$ & $-94\pm59$ &  $0.24$ & $2.83$\\
RRcs4  & $-8.97E-4\pm1.87E-4$ & $-106\pm61$ &  $0.30$ & $2.57$\\
RRcs5  & $-1.09E-3\pm1.87E-4$ & $-194\pm70$ & $0.22$ & $3.17$\\
RRcs6  & $-3.06E-4\pm1.88E-4$ & $-106\pm73$ & $0.25$ & $2.59$\\
RRcs7  & $-9.10E-4\pm1.19E-4$ & $-70\pm39$ &  $0.39$ & $4.92$\\

RRcs8  & $-8.97E-4\pm1.87E-4$ & $-86\pm78$ &  $0.59$ & $3.39$\\
RRcs9  & $-3.88E-4\pm1.87E-4$ & $-92\pm62$ &  $0.33$ & $4.87$\\
RRcs10  & $-5.59E-4\pm1.66E-4$ & $-165\pm59$ &  $0.40$ & $1.89$\\
\hline
RRcl$1_1$  & $-1.22E-3\pm7.41E-5$ & $-137\pm22$ &  $0.15$ & $3.04$\\
RRcl$1_2$  & $-1.14E-3\pm9.90E-5$ & $-70\pm25$ &   $0.20$ & $4.09$\\
RRcl$2$  & $-1.16E-3\pm4.38E-5$ &  $-166\pm32$ &   $0.37$ & $3.69$\\
RRcl$3_1$  & $6.11E-4\pm2.85E-5$ &  $-56\pm13$ &   $0.26$ & $4.76$\\
RRcl$3_2$  & $6.08E-4\pm4.50E-5$ & $-42\pm16$ & $0.26$ & $3.72$\\
RRcl$4$  & $-1.64E-4\pm6.34E-5$ &  $-212\pm31$ &  $0.13$ & $2.91$\\
RRcl$5$  & $-1.27E-3\pm6.28E-5$ &  $-183\pm31$ &  $0.34$ & $5.73$\\
RRcl$6$  & $9.95E-5\pm5.82E-5$ &  $-76\pm26$ &   $0.37$ & $2.29$\\
RRcl$7$  & $1.20E-4\pm7.93E-5$ &  $-79\pm38$ &   $0.67$ & $3.07$\\
RRcl$8$  & $-8.16E-4\pm9.95E-5$ & $-150\pm31$ &  $0.18$ & $1.84$\\
RRcl$9_1$  & $-1.35E-3\pm5.77E-5$ &  $-75\pm23$ &  $0.43$ & $2.39$\\
RRcl$9_2$  & $-1.36E-3\pm9.86E-5$ & $-90\pm37$ &  $0.39$ & $2.47$\\
RRcl$10$  & $-1.73E-3\pm9.90E-5$ & $-211\pm20$ &  $0.14$ & $1.37$\\
\hline
RRdl1  & $-1.26E-3\pm2.93E-5$ & $-149\pm15$ & $0.20$ & $3.22$\\
RRdl2  & $-1.16E-3\pm3.30E-5$ & $-89\pm12$ &  $0.27$ & $5.69$\\
RRdl3  & $-1.71E-3\pm5.42E-5$ & $-222\pm20$ & $0.12$ & $1.75$\\
\hline
\hline
\end{tabular}
\end{center}
\tablecomments{\\\hspace{\textwidth}
 1. $z_{\rm e1,\alpha}$ represents the redshift of the emission component of the ``first apparition'' in the observer's frame.\\\hspace{\textwidth}
 2. $V_{\rm e1,\alpha}$ represents the radial velocity of the emission component of the ``first apparition'' in the stellar rest frame.\\\hspace{\textwidth}
 3. Flux$_{\rm e1,\alpha}$ indicates the normalized flux of the emission. \\\hspace{\textwidth}
 4. FWHM$_{\rm e1,\alpha}$ indicates full width at half maximum of the emission. \\\hspace{\textwidth}
 5. The names like RRcl$1_1$ mean that there are not only one spectrum of one star.}
\end{table*}

\vskip 0.5cm
\section{Discussion}\label{section:discussion}

Emission features in the spectra of RR Lyrae stars can be interpreted as shocks propagating through their pulsating envelopes. The brightest emission, the ``first apparition'', can be produced by a shock forming below the photosphere near the time when the star achieves a minimum radius. The shock accelerates outward and the atoms de-excite after being excited by the shock in the radiative wake. The ramp pressure at the shock front provides energy which should be high enough to excite neutral hydrogen from the second quantum state upwards. The strength or values linked to the emission and the intensity of the shock front determine whether the emission is possible or not. Thus, our discovery possibly indicates the existence of shock waves in non-fundamental mode RR Lyrae stars.

The shock wake is suggested to be relatively narrow compared to the radius of the photosphere \citep{Chadid2011Carnegie}. \cite{Chadid2014AJ....148...88C} put forward a new scenario of shock propagation, which demonstrates that the $Sh_{PM1}$, $Sh_{PM}$, $Sh_{PM2}$, and $Sh_{PM3}$ are receding in the Eulerian rest coordinates when advancing in the Lagrangian coordinates during the time interval of the ascending branch of the radial velocity curve.

We calculate the radial velocity of the emission component of the ``first apparition'' in the stellar rest frame, which is related to the physical conditions occurring in the radiative zone of the shock. The equation is as follows:
\begin{eqnarray}\label{eq:Vshock}
V_{\rm e1,\alpha} = c\frac{(\lambda_{\rm e1,\alpha}-\lambda_{\rm ab})}{\lambda_{0}},
\end{eqnarray}
where $\lambda_{\rm e1,\alpha}$ indicates the wavelength corresponding to the central wavelength of the emission component. $\lambda_{\rm ab}$ represents the central wavelength of the absorption component. $\lambda_{0}$ is the laboratory wavelength.

The emission features in our sample spectra are rather moderate with respect to the continuum. Some of the measurement results of $V_{\rm e1,\alpha}$ exceed 100 km/s. At such a velocity, the emission of hydrogen would be far above the continuum if the exciting shock front were located in the photospheres. More studies on the propagation of radiation are needed at this point.

The possible detection of shock waves in the atmosphere of RR Lyrae also influences the understanding of the origin of the Blazhko effect.
The Blazhko effect is a feature frequently observed on the light curves of RRab and RRc stars. 
According to \cite{Chadid2014AJ....148...88C}, this effect is a complicated dynamical process due to the interactions of multi-shock and the outflowing wind in stellar coronae. Blazhko RRab stars normally resemble RRc stars of minimum Blazhko amplitude, but it cannot be confirmed that they share the same physical mechanism. The origins of the Blazhko effect in RRab and RRc stars were suspected to be totally different due to the absence of detection of shock waves in RRc stars.  
Unfortunately, the small number of observed Blazhko RRc stars limits our knowledge of their embedded physics. 
The large variety of resonant, nonresonant, and chaotic possible states should be taken into considerations to explain low-amplitude variations \citep{Moln2012ApJ...757L..13M,Moln2012AN....333..950M}.  
 In this sense, our research will serve as a valuable constraint on the future investigation of long-term modulations in non-fundamental mode stars and provide a clue to the complicated shock mechanism.

\section{Analysis of frequency components in RRdl3}\label{section:RRdl3}

According to the photometric observations from ZTF DR2, we notice that the light curve of one of the multi-mode RR Lyrae stars with the ``first apparition'', RRdl3, may feature Blazhko-type modulation. Blazhko-type modulation in four double-mode RR Lyrae stars has been discovered in the globular cluster M3 by \cite{Jurcsik2014ApJ...797L...3J}. \cite{Smolec2015MNRAS.447.3756S} also reported the discovery of 15 RRd stars in the Optical Gravitational Lensing Experiment (OGLE) Galactic bulge collection.

We analyze $g$-band light curve of RRdl3 with a timespan of 394.07 days, using a standard successive pre-whitening technique \citep{Moskalik2009MNRAS.394.1649M}. Outliers have been removed. At each step we fit the data using a non-linear least-square procedure with the sine series of the following form:
\begin{eqnarray}\label{eq:sineseries}
m(t) = m_0 + \sum^N_{k=1} A_k {\rm sin}(2\pi f_k t+\phi_k),
\end{eqnarray}
where $f_k$ are independent frequencies detected in the discrete Fourier transform of the data and their possible linear combinations. The residuals of the fit are used to search for frequencies in the next step. Then, a new Fourier series consisting of all frequencies detected so far are fitted to the data again. The process continues until no new significant frequency is detected and the residuals are virtually white noise. The frequencies, e.g., $f_a$ and $f_b$, are regarded as unresolved if $\left|f_a-f_b\right|<2/T$, where $T$ represents the length of the observed dataset. Here $2/T\approx0.005$. 

\begin{figure*}[tb]
\begin{center}
\begin{tabular}{c}
\includegraphics*[width=180mm,height=16.2cm]{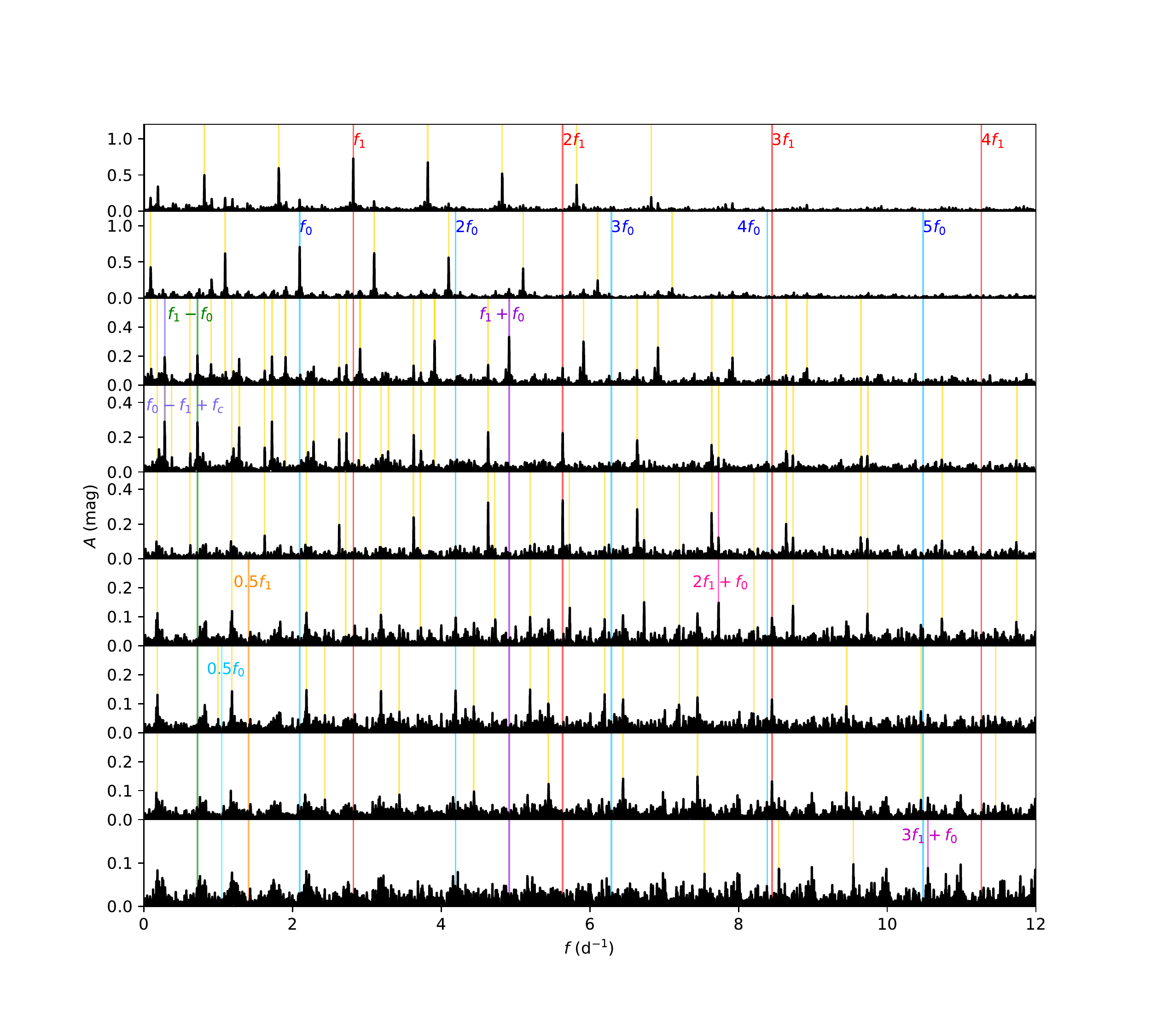}
\end{tabular}
\caption{\label{fig:prewhitening} Pre-whitening sequence for RRdl3. Uppermost panel shows power spectrum of original data. Lower panels display power spectra after removing consecutive frequencies. Feature lines are marked with different colors. Yellow lines are linear combinations expressed by $\lambda_0f_0+\lambda_1f_1+\lambda_cf_c$, while $f_c$ is the daily cadence.}
\end{center}
\end{figure*}

The pre-whitening sequence for RRdl3 is displayed in Figure~\ref{fig:prewhitening}. Apart from $f_0$, $f_1$ and their linear combinations $\lambda_0f_0+\lambda_1f_1$, we detect many significant signals of frequencies that can be expressed as $\lambda_0f_0+\lambda_1f_1+\lambda_cf_c$, where $f_c\approx1.003$. They manifest in the spectrum of frequency as equally spaced multiplets around main frequencies and their harmonics. Obviously, $f_c$ is generated from the daily cadence, which is a major problem for data analyses in ground-based time-domain surveys (see Section 5.3 in \cite{Chen2020}). In the pre-whitening process, after we removed the main peak $\lambda_0f_0+\lambda_1f_1$, the side peaks $\lambda_0f_0+\lambda_1f_1+\lambda_cf_c$ disappear as well.

\begin{table}[tbp]
\caption{\label{table:RRdl3} Basic properties of RRdl3 derived from the analysis of the ZTF photometry.}
\begin{center}
\begin{tabular}{lccccc}
\hline
\hline
{Object}   & $P_0({\rm d})$ & $P_1({\rm d})$ & $P_1/P_0$ & $A_0(\rm mag)$ & $A_1(\rm mag)$ \\
\hline
RRdl3  & $0.47701$ & $\textbf{0.35501}$ & $0.74424$ & $0.125$ & $ 0.237$  \\

\hline
\hline
\end{tabular}
\end{center}
\tablecomments{Period of the dominant mode is highlighted with bold font.}
\end{table}

\begin{figure*}[tb]
\begin{center}
\begin{tabular}{cc}
\includegraphics*[width=90.3mm,height=6cm]{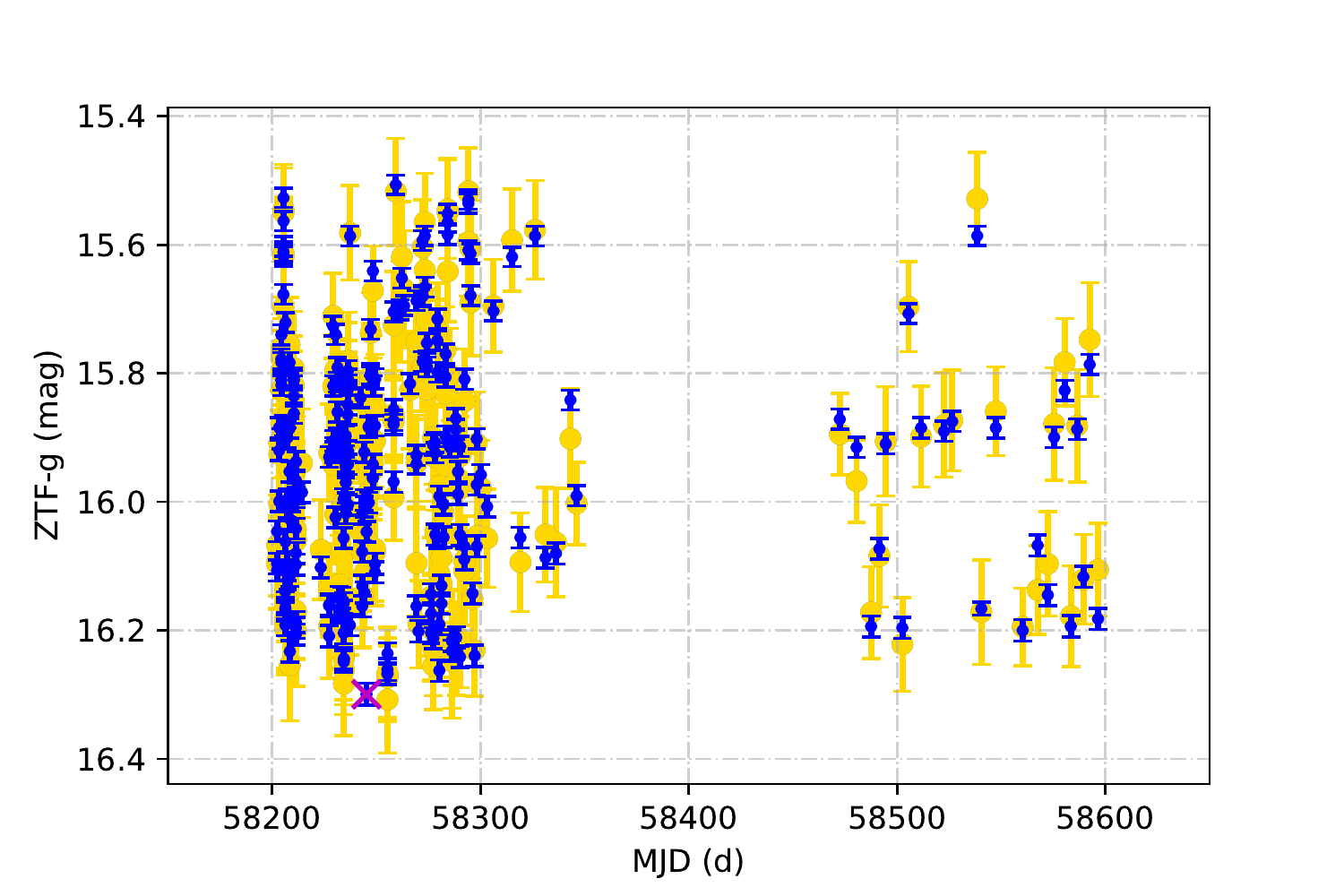}&
\includegraphics*[width=90.3mm,height=6cm]{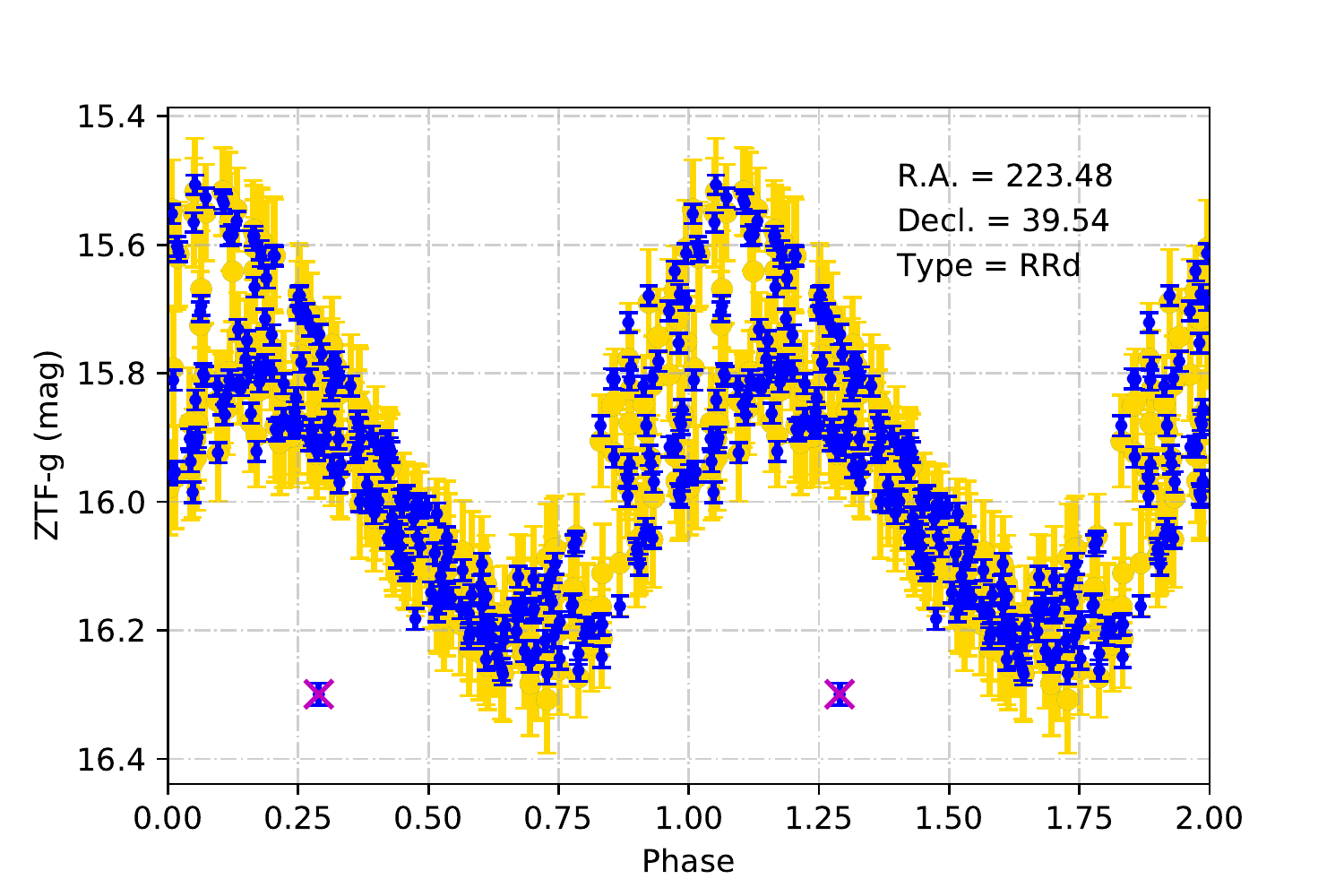}
\end{tabular}
\caption{\label{fig:newlightcurve} Regeneration of light curve of RRdl3 with full light curve solution. Blue points with error bars are observations from ZTF-g. Outliers are removed. Yellow points with error bars are regenerated points with full light curve solution with the same observing time as observations. Left (a): Regenerated light curve. Right (b): Folded regenerated light curve using the first overtone period measured from observations.}
\end{center}
\end{figure*}

We find a $full$ $light$ $curve$ $solution$ for RRdl3. Basic properties of RRdl3 derived from the analysis of ZTF-$g$ are displayed in Table~\ref{table:RRdl3}. 
The period of the dominant mode is highlighted with bold font. 
According to the resolution calculated by data length, the last two digits of periods are given only for reference. 
The full light curve solution is shown in Table~\ref{table:fulllightcurvesolution}. In order to avoid overfitting, $f_1-f_0+f_c$ and $2f_1+f_0+f_c$ are used instead of $f_1-f_0$ and $2f_1+f_0$ according to the process of pre-whitening and fitting. The regenerated light curves are displayed in Figure~\ref{fig:newlightcurve}. 

\begin{table}[tbp]
\caption{\label{table:fulllightcurvesolution} Full light curve solution for RRdl3.}
\begin{center}
\begin{tabular}{lcccccc}
\hline
\hline
{freq.id}   & $f(\rm d^{-1})$ & $A(\rm mag)$ & $\sigma$ & $\phi(\rm rad)$ & $\sigma$ & sig \\
\hline
$f_1-f_0$  & $0.72044$ &    &    &    &   &   \\
$f_c$  & $1.00300$ &  &  &  &  &  \\
$f_1-f_0+f_c$  & $1.72344$ & $0.037$ & $0.002$ & $1.089$ & $0.064$ & $30.314$  \\
$f_0$  & $2.09641$ & $0.125$ & $0.002$ & $1.551$ & $0.018$ & $93.020$  \\
$f_1$  & $2.81685$ &  $0.238$ & $0.002$ & $3.847$ & $0.010$ & $147.464$  \\
$2f_0$  & $4.19282$ & $0.016$ & $0.002$ & $5.337$ & $0.140$ & $13.802$ \\
$f_1+f_0$  & $4.91326$ & $0.047$ & $0.002$ & $1.607$ & $0.050$  & $31.349$ \\
$2f_1$  & $5.63370$ & $0.030$ & $0.002$ & $4.701$ & $0.077$ & $30.614$  \\
$2f_1+f_0$  & $7.73011$ &    &    &    &  & $ $    \\
$3f_1$  & $8.45055$ & $0.015$ & $0.002$ & $5.452$ & $0.162$ & $14.276$  \\
$2f_1+f_0+f_c$  & $8.73311$ & $0.017$ & $0.002$ & $1.428$ & $0.135$ & $13.998$  \\
\hline
\hline
\end{tabular}
\end{center}
\tablecomments{Consecutive columns are frequency id., frequency value, amplitude with standard error, and phase with standard error. The table is sorted by increasing frequency. According to the resolution calculated by data length, the last two digits of frequencies are given only for reference. 
In order to avoid overfitting, $f_1-f_0+f_c$ and $2f_1+f_0+f_c$ are used instead of $f_1-f_0$ and $2f_1+f_0$ according to the process of pre-whitening and fitting.}
\end{table}

\begin{figure*}[tb]
\begin{center}
\begin{tabular}{c}
\includegraphics*[width=180mm,height=10cm]{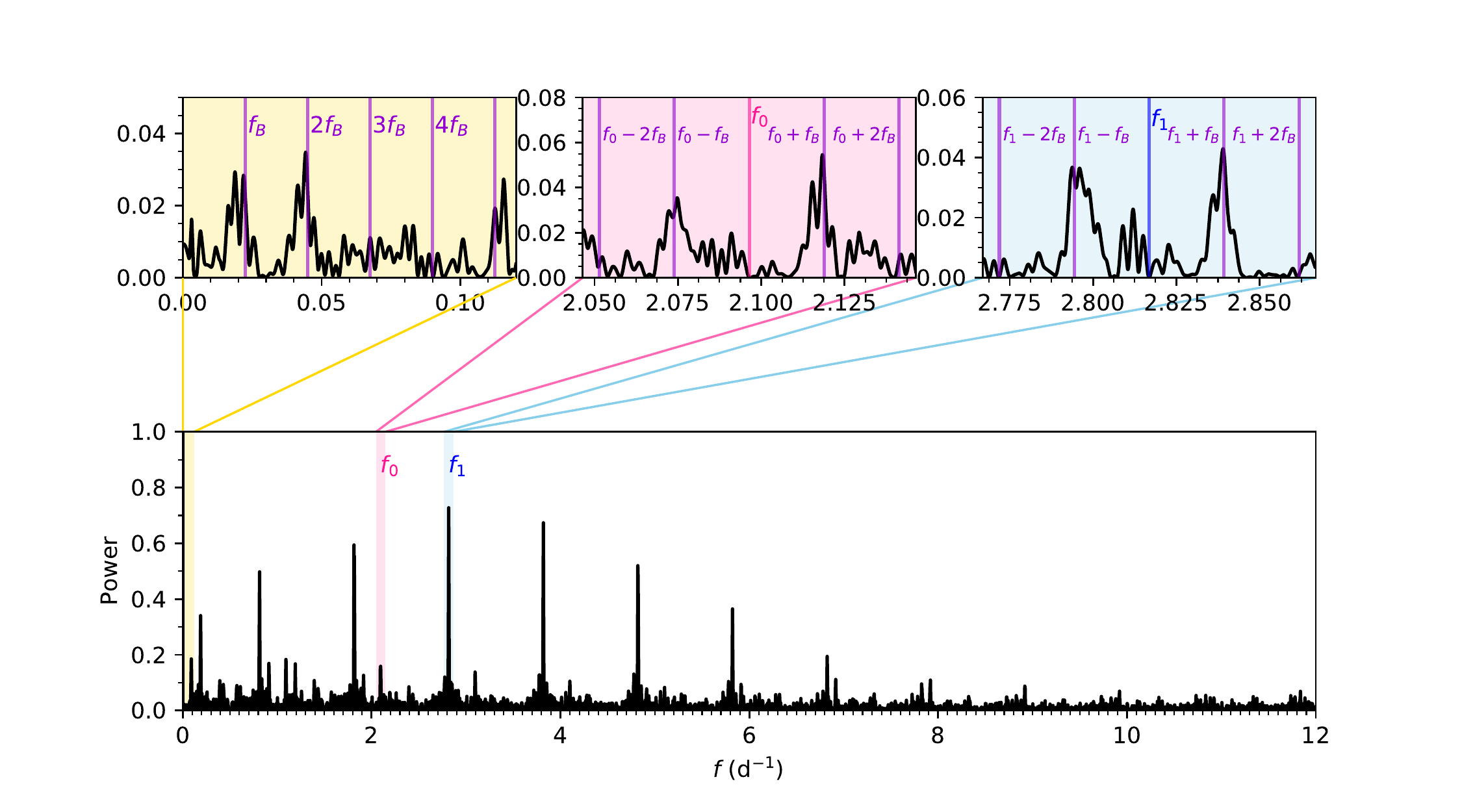}
\end{tabular}
\caption{\label{fig:Blazoom} Pre-whitened power spectrum of RRdl3. Lower panel shows power spectrum of original data. Upper panels show power spectra after pre-whitening, displaying the fine structures in low frequency, around $f_0$, and around $f_1$, respectively.}
\end{center}
\end{figure*}

The RR Lyrae stars which show additional close side peaks at the fundamental and/or the first overtone frequency are suspected of long-term modulation \citep{Smolec2015MNRAS.447.3756S}. The signals of modulation are shown as equally spaced triplets or close doublets in a typical ground-based observation \citep{Alcock2003ApJ...598..597A,Moskalik2003A&A...398..213M}. The inverse of separation on the frequency spectrum between multiplet components is the modulation period. In Figure~\ref{fig:Blazoom}, we display the amplifying pattern. We can see a series of weak signatures which indicates a possible Blazhko-type modulation, with $f_B\approx0.022$ and $P_B\approx44.483{\rm d}$. Visible signals are seen at $f_0-f_B$, $f_0+f_B$, $f_1-f_B$, and $f_1+f_B$. But the signals are not enough significant compared to others, probably because it is hampered by a lack of quantitative data. A larger and more homogenous photometric dataset is required for a clearer Blazhko-type modulation analysis.

\section{Conclusions}\label{section:conclusion}

We report the first systematic search for Balmer emission features in first-overtone and multi-mode RR Lyrae stars, taking advantage of large spectral surveys. In this work, we discover 23 cases in total, including ten RRc stars in SDSS, ten RRc in LAMOST, and three RRd stars in LAMOST. The basic parameters and measurements of the properties of the ``first apparitions'' are displayed in Table~\ref{table:parameter},  and \ref{table:measurement}.

The targets are selected through our handcrafted one-dimensional pattern recognition pipeline, using low-resolution single-epoch spectra. 
We fit the ``first apparition'' using two $S\acute{e}rsic$ profiles, and estimate uncertainties by error propagation for the covariance matrix and Monte Carlo method. We calculate the radial velocities of the emission lines, which are related to the physical conditions occurring in the radiative zone of the shock in which the hydrogen emission is formed.  
Moreover, with photometric observations from ZTF DR2, we present a detailed analysis of the light curve of RRdl3. We find a $full$ $light$ $curve$ $solution$ for RRdl3. 
The result suggests that a series of weak signatures possibly indicate the period of its Blazhko-type modulation. 
To draw a solid conclusion on this point, a larger and more homogenous photometric dataset is still required for a more precise analysis.

The detection of hydrogen emission lines in the first-overtone and multi-mode RR Lyrae variables indicates the possible existence of shock waves, which gives us a new insight into the origin of the Blazhko effect. With further observational evidence of shock wave signals in non-fundamental mode RR Lyrae stars, we will finally unveil the role of shock waves in the long-term modulation of them.

\vskip 1cm






\section*{Acknowledgements}
The suggestions and comments by anonymous referees are gratefully acknowledged. We appreciate the help from Dr. Hao-Tong Zhang, Dr. Zhong-Rui Bai, and Dr. Jian-Jun Chen for accessing the single-epoch spectra from LAMOST. We acknowledge research support from the Cultivation Project for LAMOST Scientific Payoff and Research Achievement of CAMS-CAS. Li-Cai Deng acknowledges research support from the National Science Foundation of China through grants 11633005. Xiao-Dian Chen also acknowledges support from the National Natural Science Foundation of China through grant 11903045. Xiao-Wei Duan acknowledges support from the Peking University President Scholarship. Hua-Wei Zhang acknowledges research support from the National Natural Science Foundation of China (NSFC) under No. 11973001 and National Key R$\&$D Program of China No. 2019YFA0405504. We thank Sarah A. Bird, Wei-Jia Sun, Gregory Joseph Herczeg, Meng Zhang, Xing-Yu Zhou, and Hassen Yesuf for discussing and language editing of the paper. We thank Mark Taylor for the TOPCAT software. 
The CSS survey is funded by the National Aeronautics and Space Administration under Grant No. NNG05GF22G issued through the Science Mission Directorate Near-Earth Objects Observations Program.  The CRTS survey is supported by the U.S.~National Science Foundation under grants AST-0909182 and AST-1313422. 
This work has made use of data from the European Space Agency (ESA) mission {\it Gaia} (\href{https://www.cosmos.esa.int/gaia}{https://www.cosmos.esa.int/gaia}), processed by the {\it Gaia} Data Processing and Analysis Consortium (DPAC, \href{https://www.cosmos.esa.int/web/gaia/dpac/consortium}{https://www.cosmos.esa.int/web/gaia/dpac/consortium}). Funding for the DPAC has been provided by national institutions, in particular the institutions participating in the {\it Gaia} Multilateral Agreement. 
The Gaia archive website is \href{https://archives.esac.esa.int/gaia}{https://archives.esac.esa.int/gaia}. 
This publication makes use of data products from the Wide-field Infrared Survey Explorer, which is funded by the National Aeronautics and Space Administration. The All-Sky Automated Survey for Supernovae (ASAS-SN) group are partially funded by Gordon and Betty Moore Foundation 5-year grant GBMF5490. They are also supported by NSF Grants AST-151592 and AST-1908570. ATLAS construction and operations are funded by grants 80NSSC18K0284 and 80NSSC18K1575 under the NASA Planetary Defense Office and Near-Earth Objects Observations program (NEOO). 
This work also made use of data from the Zwicky Transient Facility project (ZTF), based on observations obtained with the Samuel Oschin 48-inch Telescope at the Palomar Observatory as part of the Zwicky Transient Facility project. 
ZTF is supported by the National Science Foundation under Grant No. AST-1440341. 
Funding for SDSS-III has been provided by the Alfred P. Sloan Foundation, the Participating Institutions, the National Science Foundation, and the U.S. Department of Energy Office of Science. The SDSS-III web site is \href{http://www.sdss3.org/}{http://www.sdss3.org/}.  
 The Guoshoujing Telescope (the Large Sky Area Multi-Object Fiber Spectroscopic Telescope LAMOST) is a National Major Scientific Project built by the Chinese Academy of Sciences. Funding for the project has been provided by the National Development and Reform Commission. LAMOST is operated and managed by the National Astronomical Observatories, Chinese Academy of Sciences.

\vspace{5mm} 
\software{NumPy \citep{NumPyoliphant2006guide}, SciPy \citep{Virtanen2020SciPy-NMeth}, AstroPy 
\citep{Astropyprice2018astropy}, Matplotlib \citep{Matplotlibhunter2007matplotlib}, 
Scikit-learn \citep{Scikitpedregosa2011scikit}, 
PyAstronomy \citep{Czesla2019ascl.soft06010C}, TOPCAT \citep{Taylor2005ASPC..347...29T}}

\bibliographystyle{aasjournal}  
\bibliography{mainDXW}



\end{document}